\documentclass[11pt]{cernrep}
\usepackage{graphicx}
\usepackage{here}
\def\brlist{\begin{thebibliography}{99}}
\def\brf{\bibitem}
\def\erlist{\end{thebibliography}}
\def\mcite{$\,$\cite}

\begin{document}
\hyphenation {che-ren-kov}
\hyphenation {non-per-tur-ba-tive}
\hyphenation {per-tur-ba-tive}
\title{HYBRID MESON PRODUCTION VIA PION SCATTERING 
FROM THE NUCLEAR COULOMB FIELD}
\author{Murray Moinester}
\institute{R. and B. Sackler Faculty of Exact Sciences, 
School of Physics and Astronomy, Tel
Aviv University, 69978 Tel Aviv, Israel;
e-mail: murraym@tauphy.tau.ac.il}
\maketitle

\begin{abstract}

The CERN COMPASS experiment can use 100-280 GeV pion and kaon beams and magnetic
spectrometers and calorimeters to measure Hybrid (mixed quark-gluon) meson
production cross sections in the Primakoff scattering of high energy pions and
kaons from virtual photons in the Coulomb field of High-Z targets. There are many
advantages to studying such processes via the Primakoff reactions $\pi^- \gamma
\rightarrow Hybrid \rightarrow \rho\pi, \eta\pi, \eta'\pi, \pi b_1, \pi f_1$, and
similar reactions with a K$^-$ beam. Such data should provide significant input
for gaining a better understanding of non-perturbative QCD. A brief description
and update of this program is presented.

\end{abstract}

\section {PIONIC HYBRID MESONS}

The CERN COMPASS experiment focuses on issues in the physics of strong
interactions pertaining to the structure of hadrons in terms of valence quarks
and gluons. This report deals with the COMPASS Hybrid meson program. The Hybrid
(quark-antiquark-gluon, $q\bar{q}g$) mixed quark-gluon mesons are particles
predicted by QCD \cite {jlabhybrid,azas,sgjn,suc}. The unambiguous discoveries of
these meson states will provide a major landmark in hadron spectroscopy. Of
course, the Fock state of a meson is represented as an infinite expansion of
different quark and gluon configurations, but in a Hybrid meson, the $q \bar {q}
g$ component dominates. The force between quarks in QCD is mediated by the
exchange of colored gluons. In ordinary quark-antiquark mesons, the exchanged
gluons localize within a narrow string-like tube connecting the quarks: a
non-vibrating fluxtube of colored flux lines. In the fluxtube description of a
Hybrid meson, a quark-antiquark pair couples directly to the vibrational degrees
of freedom of the gluonic fluxtube. Hybrid mesons therefore contain explicit
valence gluons, as opposed to the hidden gluons in ordinary mesons \cite {sgjn}.
Understanding such valence gluons is critical to understanding the origin of
hadron mass. Establishing the existence of Hybrid mesons and studying their
properties can provide insight into color confinement.

Input from experiments is needed to provide better understanding of the current
situation, especially to show that the present evidence \cite {dunn} is not just
the result of some artifice. GSI Darmstadt \cite {GSI} proposes an exotic meson
program via a new facility for $p \bar{p}$ reactions. The planned \$150M 12 GeV
JLab upgrade focuses on Hybrid meson studies at JLab Hall D \cite {HallD}. Future
experiments at Fermilab CDF and D0 may measure double-Pomeron production of
exotic mesons \cite {chpc}. More results from BNL, VES, Crystal Barrel \cite
{cb,E852,whs} (from analysis of completed experiments), and further theoretical
calculations, are becoming available. COMPASS has a multi-faceted Hybrid meson
research program \cite{don,dor}, which includes this proposal for Primakoff
production of Hybrids. COMPASS \cite {compass} is well positioned to collect
significantly cleaner data, and at lower cost, and using complementary methods,
compared to the other planned efforts. The Primakoff production of Hybrids in
COMPASS is part of a more global Primakoff physics program, involving studies of
pion and kaon polarizabilities, and studies of the chiral anomaly \cite
{compass,bormio}.

The COMPASS Hybrid meson searches \cite {bormio,hybrid,tm} will focus on
`oddballs'---mesons with quantum numbers not allowed for ordinary $q\bar{q}$
states, such as $I^{G}~J^{PC} = 1^{-}~1^{-+}$ Hybrids. Unlike Glueballs, these
cannot mix with normal $q\bar{q}$ mesons. However, since such oddballs could also
be four-quark states, the spin assignment is not sufficient to make the Hybrid
identification \cite {suc2}. In certain cases, there are other theoretical
interpretations \cite {dp,sb} for such experimental signals. Previous
experimental efforts have reported several 1$^{-+}$ resonant signals \cite
{cb,E852,whs} at masses between 1.4 and 1.9 GeV, in a variety of decay channels,
including the $\rho\pi$ channel. The signature for such an exotic state is that a
detailed partial wave analysis (PWA) of a large data sample requires a set of
quantum numbers inconsistent with a normal ($q \bar{q}$) meson. To increase
confidence in a Hybrid interpretation, complementary evidence is sorely needed.
The path to understanding requires that different experiments, such as the
important input from COMPASS Primakoff studies, provide a large database of
candidate Hybrid states, and their properties.

Barnes and Isgur, using the flux-tube model \cite {tb,ni}, calculated the mass of
the lightest pionic hybrid with quantum numbers of $J^{PC}=1^{-+}$ to be around
1.9 GeV, higher than the experimental claims. Close and Page \mcite{fluxtb}
predict that such a high-lying pionic hybrid should decay into the following
channels:

\begin{center}\begin{tabular}{c|c|c|c|c}
   $b_1\pi$&$f_1\pi$&$\rho\pi$&$\eta\pi$&$\eta'\pi$\\
\hline
170&60&$5\to 20$&$0\to 10$&$0\to 10$\\
\end{tabular}\end{center}

\noindent where the numbers refer to the partial widths in MeV. They expect the
total width to be larger than 235-270 MeV, since the $s \bar{s}$ decay modes were
not included. Recent updates on hybrid meson structure are given in Refs.
\cite{sgjn,nib98,kpb98}. The lower lying Hybrids of recent experiments have
significantly different branching ratios. This may reflect that the structure of
these experimental low-mass Hybrids is different than that of the theoretical
high-mass Hybrids, and considering the high thresholds of the $b_1\pi$ and
$f_1\pi$ decay channels for decay of a low-mass Hybrid.

   From over a decade of experimental efforts at IHEP \cite {ihep1,ihep2,ves},
CERN \cite {cb,na12}, KEK \cite{kek}, and BNL \cite {E852}, several hybrid
candidates have been identified. BNL E852 \cite {E852} reported two $J^{PC}=
1^{-+}$ resonant signals at masses of 1.4 and 1.6 GeV in $\eta\pi^-$ and
$\eta\pi^0$ systems, as well as in $\pi^+ \pi^- \pi^-$, $\pi^- \pi^0 \pi^0$,
$\eta' \pi^-$ and $f_1(1285) \pi^-$. The VES collaboration presented \cite {whs}
the results of a coupled-channel analysis of the $\pi 1(1600)$ meson in the
channels $b_1(1235)\pi$, $\eta'\pi$, and $\rho\pi$, with a total width of 290
MeV, and relative branching ratios: $1:1\pm0.3:1.6\pm0.4$. They did not include
the $f_1 \pi$ channel in this analysis. The resonant nature of the 1.6 GeV state
was observed in the $b_1 \pi$ mode, by a combined fit of the 2$^{++}$ and
1$^{-+}$ waves. Then an assumption was made that in the $\eta' \pi$ and $\rho
\pi$ channels, they observed the same state, considering the similar shapes.
Therefore, their coupled channel analysis result of a $\rho \pi$ partial width of
130 MeV, and a total width of 290 MeV, should be taken with some caution,
considering also that the $f_1 \pi$ channel is not included. Still, for the 1.6
GeV region, the VES $\rho \pi$ partial width is consistent with the BNL $\rho
\pi$ width of 168 MeV. For count rate estimates below, we will use an average
value of 150 MeV for the $\rho \pi$ width at 1.6 GeV.

\begin{figure}
\begin{center}
\includegraphics[width=7cm]{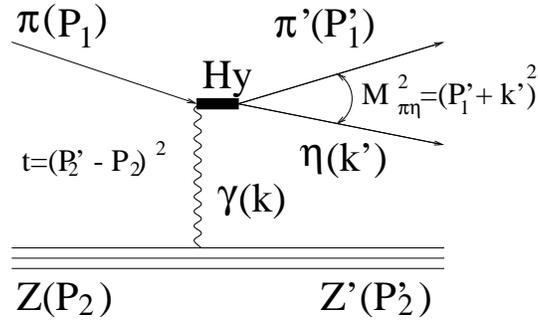}
\caption{Primakoff Hybrid-meson  production process and
kinematic variables (4-momenta): P1, P1$^\prime$ = for initial/final pion,
P2, P2$^\prime$ = for initial/final target, k = for initial $\gamma$,
k$^\prime$ = for final $\eta$.}
\label{fig:diagram}
\end{center}
\end{figure}

The kinematic variables for the $\pi \gamma \rightarrow HY \rightarrow \pi^-
\eta$ Primakoff process in COMPASS are shown in Fig.~\ref{fig:diagram}. A
virtual photon from the Coulomb field of the target nucleus interacts with the
pion beam. At 200 GeV $\pi$ beam energy, nuclear (meson-exchange) amplitudes in
the Primakoff production of the $\rho$ meson were shown to be very small \cite
{jens} in the kinematic region of Primakoff production. This is very
encouraging for the COMPASS studies. Still, COMPASS can carry out data
analysis of Hybrid data with and without meson-exchange amplitudes, to
test to what extent their presence effects the extraction of the
Primakoff cross section. For a given Hybrid ($\pi 1$) partial decay width
$\Gamma(\pi 1 \rightarrow \pi \rho$), Vector Dominance Model (VDM) gives an
associated radiative width $\Gamma(\pi 1 \rightarrow \pi \gamma$) \cite
{zihy,tf}. As the Primakoff cross section is proportional to $\Gamma(\pi 1
\rightarrow \pi \gamma$) \cite{zihy,tf,mol}, all Hybrids that decay to the $\pi
\rho$ channel should also be produced in the Primakoff reaction. In contrast to
the $\pi 1(1600)$, the $\pi 1 (1400)$ 1$^{-+}$ state should be only weakly
populated in the Primakoff reaction, as it has a very small $\pi \rho$ partial
decay width. For the $\pi 1$(1600) radiative width, we use the standard VDM
expression \cite{zihy,tf,jens} with a $\rho\gamma$ coupling $g^2_{\rho
\gamma}/\pi=2.5$, which correctly relates the corresponding widths for the
$\rho$ meson decay. We obtain $\Gamma(\pi 1 \rightarrow \pi \gamma) = 6.6
\times 10^{-3} \Gamma(\pi 1 \rightarrow \pi \rho)$, which gives the VDM
estimate $\Gamma (\pi 1 \rightarrow \pi \gamma) \approx 1000 KeV$.

In Fig. 1, a Hybrid meson (other than $\pi 1 (1400)$) is produced and decays to
$\pi^- \eta$ at small forward angles in the laboratory frame, while the target
nucleus (in the ground state) recoils coherently with a small transverse p$_T$.
The corresponding small p$_T$ of the exchanged photon means that it is
essentially real and transverse. Consequently, the helicities of incident and
Primakoff produced mesons differ by unity. The peak at small target $p_T$ used to
identify \cite {zihy,tf} the Primakoff process is observed by using the beam and
vertex and other COMPASS detectors to measure the beam pion and final state
Hybrid momenta. Primakoff scattering is a large impact-parameter,
ultra-peripheral reaction on a virtual photon target. The initial state pion and
final state Hybrid interact at very small t-value (four momentum transfer to the
target nucleus), where the nuclear form factor is essentially unity, and there
are no final state interactions \cite {future}.

  It is important to note that in addition to Primakoff production of
excited meson states, one also expects diffractive production. It is
necessary to separate the Primakoff forward peak from these underlying
sometimes large diffractive yields \cite{zihy,mol}. The strength of the
diffractive cross section depends on angular momentum conservation at zero
degrees; and therefore on which decay mode is studied: $b_1 \pi$, $f_1
\pi$, $\rho\pi$, $\eta\pi$, $\eta'\pi$. The potentially large
diffractively produced waves with spin projection M=0 and J$^P=1^+$ can
complicate the PWA of $\rho\pi$, but not $\pi\eta$. Of course, although
the $\pi\eta$ channel may be cleaner, it may not necessarily have a large
1$^-$ Hybrid contribution.

The t-distribution of the diffractive yield depends on whether there is
helicity flip at zero degrees. The t-dependence of these dominant (no
helicity-flip) diffractive waves is exp(-bt). The accuracy with which one
can subtract this background depends on the fraction of the projective
helicity-flip ($M=\pm 1$) diffractive waves associated with 2$^+$ states,
which have a t-dependence $t \times exp(-bt)$. All the waves have phases
and can interfere, so that subtracting only an exponential would ignore
these effects. However, as the Coulomb waves have Gottfried-Jackson
helicity $M=\pm 1$, while strong production occurs dominantly with M=0,
the interference should be zero to lowest order \cite {diff}.

Such subtraction was done, even for a large diffractive cross
section \cite {diff}, as in the recent measurement of the radiative decay
width of the a2(1320) meson \cite{mol} via its decay to the $\pi\rho$
channel. The subtraction was done by extrapolation of the data away from
the forward Primakoff peak, assuming an exp(-bt) dependence of the
diffractive yields. Such theoretical assumptions on the t-dependence of
the background should have small effect on narrow resonances (104 MeV
width) such as the a$_2$. This appears to be the case, considering the
consistency of the a$_2$ radiative widths determined via Primakoff
production, and then decay via the $\pi \rho$ and $\pi \eta$ channels
\cite {mol}. But an exponential subtraction may cause larger uncertainties
for Hybrids with large (300-500 MeV) widths. The Hybrid data analysis
therefore should test the sensitivity of the results to such exponential
background assumptions.

In the recent Primakoff study of the a$_2$ radiative width by decay to $\pi\rho$,
following subtraction of the diffractive yield, due in part to poor statistics,
the resulting Primakoff mass spectrum does not show significant structure above
the clearly observed a$_2$(1320). Also, for this high mass region, a PWA is not
available. Considering the subtraction and statistical uncertainties discussed
above, this does not rule out that the $\pi\rho$ width is nonetheless large in
the region of the expected 1.6 GeV Hybrid. And with improved statistics and
resolution and acceptance, and by studying all decay modes in the same
experiment, COMPASS can potentially do a significantly better Hybrid analysis.
High lying excited meson states certainly couple to the $\pi\rho$ channel, as
shown clearly in the work of photoproduction of excited mesons, observed by their
$\pi\rho$ decay mode \cite{photo}.

Consider some typical angular distributions from Primakoff scattering, such as
those for the $\pi^- \gamma \rightarrow \eta\pi$ scattering, for different values
of $\eta\pi$ invariant mass. If these are associated with the production and
decay of a $J^{PC}= 1^{-+}$ $(d \bar{u} g)$ Hybrid state, with quantum numbers
not available to $q\bar{q}$ mesons, then a detailed partial wave analysis (PWA)
of a large data sample would indicate the need for these quantum numbers. The
partial-wave analysis (PWA) of systems such as $\eta\pi$ or $\eta' \pi$ in the
mass region below 2 GeV requires high statistics and minimum background. This
region is dominated by strong resonances (e.g., $a_2$(1320) near the 1.4 GeV
Hybrid candidate and $\pi_2$(1670) near the 1.6 GeV Hybrid candidate), and the
PWA can yield ambiguous results \cite {ihep2} for the weaker $1^{-+}$ wave. For
the Primakoff yield, following subtraction of the diffractive yield, the cross
section for 1.6 GeV hybrid production ($\pi 1$), in all decay channels, should be
larger than the background. This is so because $\Gamma(\pi_2 \rightarrow \pi
\gamma) \approx 300$~ke$\!$V, smaller than the expected $\Gamma(\pi 1 \rightarrow
\pi \gamma)\approx 1000$~ke$\!$V. In contrast, the BNL experiment \cite {E852}
for this decay channel has a background from $\pi_2$(1670) some 10 times stronger
than the Hybrid signal. Primakoff data with reduced backgrounds should
significantly diminish uncertainties in the partial wave analysis compared to the
non-Primakoff production experiments. It might allow the Hybrid to be observed in
the mass spectrum in some of the decay modes, not only via PWA. Furthermore, in
Primakoff (photon-exchange) production experiments, besides the absence of final
state interactions, meson-exchange and diffractive backgrounds can be largely
eliminated. Most of these backgrounds occur at larger values of
the four-momentum transfer t, and are easily removed by an analysis cut on the
t-value. The part of these backgrounds that extends to small-t can be largely
removed by extrapolation to small-t of the large-t data. These are important
advantages compared to previous $\pi^-p \rightarrow Hybrid$ production
experiments \cite {hybrid,tm}.

Preliminary low statistics Hybrid production data \cite {zihy,tf} at Fermilab
E272 via Primakoff scattering provide encouraging initial results for production
of a 1.6 GeV 1$^{-+}$ Hybrid. The data for decay to the $\pi f_1$ channel shows
signs of Primakoff enhancement (excess of 25 events) at small-t. This gives
$\Gamma(\pi 1 \rightarrow \pi \gamma) \times B(\pi 1 \rightarrow \pi f_1) \approx
250 KeV$. If B($\pi 1 \rightarrow \pi f_1) \approx 0.25$, that corresponds to
$\Gamma(\pi 1 \rightarrow \pi \gamma) \approx 1000 KeV$, close to the value
deduced using VDM with a 150 MeV partial width in the $\pi \rho$ channel. For
$\pi\rho$ decay of a 1.6 GeV state, FNAL E272 had insufficient statistics to
separate a Coulomb enhancement from the diffractive background. As their data was
for the combined spectra, their estimate of the upper limit to the Primakoff
Hybrid yield has large uncertainties. FNAL SELEX measured the a$_2$ radiative
decay via the $\pi \rho$ mode. The SELEX Primakoff spectrum, after subtraction of
diffractive background, shows the a$_2$ peak clearly. The E272 $\pi\rho$ data for
diffractive plus Primakoff production does not even show the a$_2$ peak. It is
difficult to use E272 data to get Primakoff limits in the $\pi\rho$ channel near
1.6 GeV. For a 150 MeV $\pi\rho$ partial width of a 1.6 GeV Hybrid, the E272 data
gives an upper limit for $\Gamma(\pi 1 \rightarrow \pi \gamma)$ close to 100 KeV.
Since this limit value has a huge error bar, the E272 estimate using the cleaner
$\pi f_1$ data is certainly more reliable. E272 also shows via the $\eta \pi$
decay mode, that $\Gamma(\pi 1 \rightarrow \pi \gamma) \times B(\pi 1 \rightarrow
\pi \eta)$ is at most 100 KeV for a $\pi$1 near 1.6 GeV, for a $\pi$1 with a
total width of 400 MeV. But the $\pi$1 branching B($\pi 1 \rightarrow \pi \eta$)
from VES and BNL is very small, so this is still consistent with $\Gamma(\pi 1
\rightarrow \pi \gamma)$ near 1000 KeV. Consequently, one may estimate from the
E272 $\pi f_1$ data that $\Gamma(\pi 1 \rightarrow \pi \gamma)$ is close to 1000
KeV. This is consistent with $\Gamma(\pi 1 \rightarrow \pi \gamma) \approx 1000
KeV$, which we obtain by applying VDM to the experimental $\Gamma(\pi 1
\rightarrow \pi \rho)$ width, as described above.

One can summarize the situation as follows. For Primakoff scattering,
Hybrid-meson production cross section depends on the strength of its $\pi\rho$
coupling, and via vector dominance to its $\pi \gamma$ coupling. Both BNL and VES
claim this decay mode for the 1.6 GeV Hybrid candidate. A low statistics $\gamma
p \rightarrow$ Meson photoproduction experiment \cite {photo} also observed
resonances (including near 1.9 GeV) in the $\pi \rho$ channel, suggesting
possible Hybrid interpretations. Based on previous data from FNAL E272 and VES,
we estimate $\Gamma(\pi 1 \rightarrow \pi \gamma) \approx 1000 KeV$ for a 1.6 GeV
Hybrid. The relevant Primakoff reactions $\pi^- \gamma \rightarrow Hybrid
\rightarrow$ $\rho\pi, \eta\pi, \eta'\pi, b_1(1235)\pi, \pi f_1$, etc., can
therefore be studied in COMPASS potentially in the 1.4-3.0 GeV mass region, which
includes all previous Hybrid candidates.

\section {KAONIC HYBRID MESONS}

  The quark content of the hybrid meson ($q \bar{q} g$) nonet should be identical
to the quark content of the regular meson ($q \bar{q}$) nonet, with identical
SU(3) decomposition in the plane of isospin I$_3$ and hypercharge Y, for the
1$^{-}$ and other spin-parity states. Thus, for every pionic $(d \bar {u}g)$
1$^{-+}$ Hybrid, there should be a flavor excited Kaonic $(s \bar {u}g)$ 1$^{-}$
Hybrid, at an excitation energy roughly 100-140 MeV higher than its pionic Hybrid
cousin, possibly narrower because of phase space. COMPASS can observe the Kaonic
Hybrids via $K^- Z \rightarrow Hybrid \rightarrow K^- \rho^0$ Z, as well as other
decay modes: $b_1 K^-$, $f_1 K^-$, $\eta K^-$, $\eta' K^-$. The backgrounds
should be low from Primakoff excitation of normal Kaonic excited mesons, as in
the case of the pionic Hybrids; and also if the Kaonic hybrids are narrower. The
first ever measurement of the Kaonic Hybrids via Primakoff scattering would be of
inherent interest, but would also provide valuable support that the analogous
pionic signals are properly identified as Hybrids. Searches for $s \bar {s} g$
Hybrids via $K^- p \rightarrow$ Hybrid have also been proposed recently \cite
{suc3}.

\section{MESON RADIATIVE TRANSITIONS}

COMPASS will also study Primakoff radiative transitions leading from the pion to
the $\rho^-$, a$_1$(1260), and a$_2$(1320), and for the kaon to K$^*$. The data
can be obtained with a particle-multiplicity trigger \cite{tm}. Theoretical
predictions for radiative transition widths are available from vector dominance
and quark models. Independent and higher precision data for these and higher
resonances would provide a useful check of the COMPASS apparatus, and would allow
a more meaningful comparison with theoretical predictions. For example, the $\rho
\rightarrow \pi \gamma$ width measurements \cite{jens,hust,capr} range from 60 to
81 keV; the a$_1$(1260) $\rightarrow \pi \gamma$ width measurement \cite{ziel} is
$0.64 \pm 0.25 $ MeV; and the a$_2$(1320) $\rightarrow \pi \gamma$ width is is
$\Gamma = 295 \pm$ 60 keV \cite{ciha} and $\Gamma = 284 \pm 25 \pm 25$ keV
\cite{mol}. For K$^* \rightarrow K \gamma$, the widths obtained previously are 48
$\pm$ 11 keV \cite {berg} and 51 $\pm$ 5 keV \cite {chan}. The above references
indicate that the formalism of Primakoff production provides an excellent
description of excited mesons, the same formalism that we use to search for
Primkoff production of Hybrids.

These as well as polarizability and chiral anomaly \cite {compass,bormio}
Primakoff measurements are important for a variety of reasons: (1) COMPASS can
significantly improve their precision, (2) COMPASS can get data for other $q \bar
{q}$ meson excited states, (3) these measurements with sufficient statistics test
our methodologies and help calibrate our apparatus for the Hybrid studies.

\section{DETAILED DESCRIPTION OF APPARATUS}

COMPASS is a fixed-target experiment that uses a 160 GeV polarized muon beam, and
pion, kaon, and proton beams. In order to achieve good energy resolution within a
wide energy range, COMPASS has a two-stage spectrometer with 1.0 Tm and 5.2 Tm
conventional magnets. The tracking stations contain different detector types to
cover a large area, and to achieve good spatial resolution in the vicinity of the
beam. Most of the tracking detectors operate on the principle of gas
amplification, while some are silicon strip detectors. At the end of each stage,
an electromagnetic and a hadronic calorimeter detects energies of photons,
electrons and hadrons. The calorimeters in the first stage and the EM calorimeter
of the second stage have holes through which the beam passes.

In COMPASS, two beam Cherenkov detectors (CEDAR), far upstream of the target,
provide $\pi/K/p$ particle identification (PID). The incoming hadron momentum is
measured in the beam spectrometer. Before and after the target, charged particles
are tracked by high resolution silicon tracking detectors. The measurement of
both initial and final state momenta provides constraints to identify the
reaction. The final-state meson momenta are measured downstream in the magnetic
spectrometer and in the $\gamma$ calorimeter. These provide a precise
determination of the p$_T$ transfer to the target nucleus, the main signature of
the Primakoff process, and the means to separate Primakoff from meson-exchange
scattering events.

We considered in detail previously the beam, target, detector, and trigger
requirements for Hybrid studies \cite {bormio,tm}. A brief description is given
below of some important components of the apparatus for Primakoff studies. The
Sept. 2002 status of the full COMPASS apparatus is described in Ref. \cite
{future}.

\subsection{Beam Requirements}

We can obtain good statistics for the pion study by using the high beam
intensities of the CERN SPS. We can take data at different beam energies and use
different targets, with both positive and negative beams, as part of efforts to
control systematic uncertainties.

  For the 120-300 GeV hadron beams, particle identification (PID) is needed to
provide pion, kaon, and proton beam tagging for positive and negative beams. For
the COMPASS beam, one expects \cite {compass} a beam intensity of 100 MHz, with
beam composition \cite {beam} roughly: 120-300 GeV/c, negative, 87-98\% pions,
7-1\% kaons, 2\%-1\% antiprotons; 120-300 GeV/c, positive, 43-2\% pions, 7-1\%
kaons, 49-97\% protons. PID is accomplished at CERN with the CEDAR detectors,
Cherenkov differential counter with achromatic ring focusing. There are two CEDAR
detectors (in series) in the COMPASS beamline \cite {bov}. They each have eight
large area PMTs arranged in a circle, preceded by a single light diaphragm (LD)
to finely fix a ring radius. A six-fold coincidence is required for the PID. The
gas pressure is varied to set the ring radius for pions or Kaons or protons at
the LD location. The narrow diaphragm mounted in CEDAR-N separates kaons from
pions up to 300 GeV/c, and can tag protons down to 12 GeV/c.

\subsection{Target and Target Detectors\label{sec:target}}

The target platform is movable and allows easy insertion of a solid target, e.g.,
a cylindrical lead plate 40 mm in diameter and 1.4 mm thick. We use silicon
tracking detectors before and immediately after the targets. These are essential
for Primakoff reactions as the angles have to be measured with a precision of
order 100 $\mu$rad. We veto target break-up events via a target recoil detector,
and by selecting low-t events in the off-line analysis.

\subsection{The $\gamma$ Calorimeter ECAL2}

The COMPASS $\gamma$ detector is equipped with 3.8 by 3.8 cm$^2$ GAMS lead-glass
for a total active area of order 2m diameter. The central area is already
completely instrumented with ADC readouts. For the precise monitoring of energy
calibration of the photon calorimeters, COMPASS will use LED and laser monitor
systems, as described in Ref. \cite{laser}. The position resolution in the second
$\gamma$ calorimeter ECAL2 for the photon is 1.0 mm, corresponding to an angular
resolution of 30 $\mu$rad. In the interesting energy range, the energy resolution
is 2-3\%. The photon acceptance is 98\% due to a beam hole of ECAL2, while the
reconstruction efficiency is 58\%, as a result of pair production within the
spectrometer.

As can be seen from Fig.~\ref{fig:diagram}, COMPASS requires reconstructed
$\eta$s for the hybrid study. The two $\gamma$s from $\eta$ decay have
half-opening angles $\theta_{\gamma\gamma}^h$ for the symmetric decays of
$\theta_{\gamma\gamma}^h= m/E_{\eta}$, where m is the mass ($\eta$) and
E$_{\eta}$ is the $\eta$ energy. (Opening angles are somewhat larger for
asymmetric decays.) In order to catch most of the decays, it is necessary to
subtend a cone with about double that angle, i.e., $\pm 2m/E_{\eta}$, neglecting
the angular spread of the original $\eta$s around the beam direction. For the
ECAL2 $\gamma$ detector, with a circular active area of 2m in diameter, the
acceptance for the $\pi\eta$ channel at 30 m from the target for $\eta$s above
E$_{\eta}$=33 GeV is therefore excellent. At half this energy, however, the
acceptance becomes quite poor. The acceptance depends of course on the Hybrid
mass, which is taken between 1.4 and 3.0 GeV for the planned COMPASS study.
Detailed Monte Carlo studies are needed for different Hybrid decay modes, for a
range of assumed masses. For the $\pi$f$_1$ channel, for example, $f_1
\rightarrow \pi\pi\eta$, the $\eta$s will have low energy, and therefore large
$\gamma$ angles. To maintain good acceptance for low energy $\eta$s, the ECAL2
diameter should be about 2m.

The available COMPASS ECAL does not have radiation hardened blocks near the beam
hole. If those will be available, it would allow increasing the beam intensity by
a factor of five, to allow substantially more statistics for the same run time.
This would clearly be a cost-effective improvement.

\subsection{The Magnetic Spectrometer and the t-Resolution}

The p$_T$ impulses of the COMPASS magnets are 0.3 GeV/c for SM1 (4 meters from
target) and 1.56 GeV/c for SM2 (16 meters from target). The fields of both
magnets are set in the same direction for maximum deflection of the beam. We
achieve good momentum resolution for the incident and final state charged and
neutral mesons, and therefore good resolution in t. The relative momentum
resolution for charged $\pi$, with all interactions accounted for, is 1\% for
energies above 35 GeV and up to 2.5\% below this mark. The angular resolution in
a single coordinate for a charged-pion of momentum p is 7.9 mrad-GeV/p. The
reconstruction efficiency for pions with energy greater than 2 GeV is 92\%.

The angular resolution for a final-state charged meson is controlled by
minimizing the multiple scattering in the targets and detectors. With a lead
target of 0.8\% interaction length (1.6 g/cm$^2$,~24\% radiation length),
multiple Coulomb scattering (MCS) of the beam and outgoing pion in the target
gives an rms angular resolution of order 32$\mu$rad, small compared to the
intrinsic tracking-detector angular resolution. The target contributes to the
resolution in the transverse momentum p$_T$ through MCS. For $t~=~p_T^2$,
including all other effects \cite{bormio,tm}, we aim for a p$_T$ resolution of
less than 15 MeV, corresponding to $\Delta t$ smaller than $\approx$ 2.5 $\times
10^{-4}$ GeV$^2$.

This resolution will provide good separation for contributions from diffractive
and meson-exchange processes. Minimum material (radiation and interaction
lengths) in COMPASS will also yield a higher acceptance, since the $\gamma$'s
will not be converted before the ECAL2, and the result is minimum $e^+e^-$
backgrounds.

\subsection {The COMPASS Primakoff Trigger}

\begin{figure}
\begin{center}
\includegraphics[width=12cm]{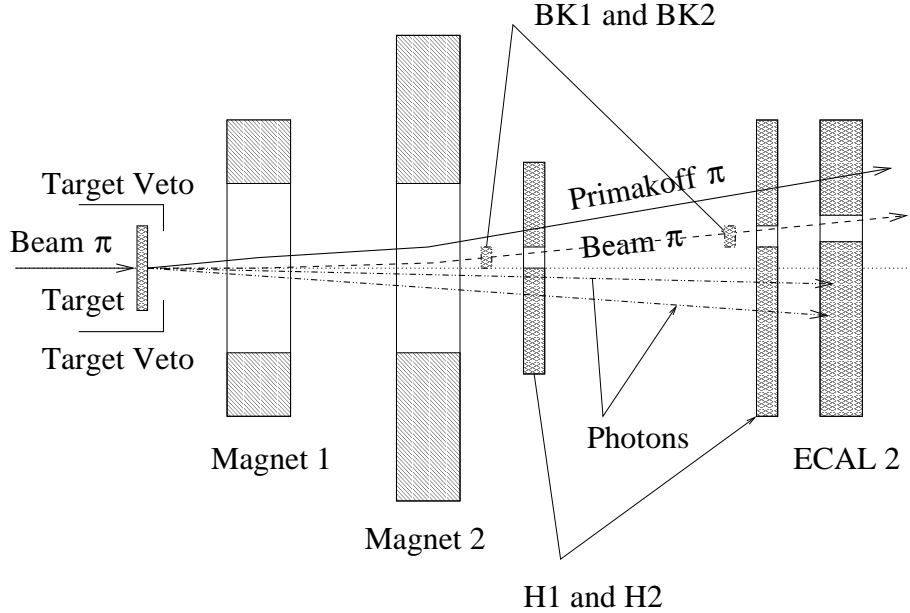}
\caption{Detector layout for the COMPASS Primakoff Hybrid trigger,
$\pi^- Z \rightarrow Hybrid \rightarrow \pi^- \eta$.
BK1,BK2=beam killer system, H1,H2=hodoscope system for charged particle 
detection, ECAL2=second photon calorimeter.
For $\eta$ decay, one observes the 
two $\gamma$s shown.  For polarizability, there is only one 
$\gamma$ to 
detect.}
\label{fig:trigger}
\end{center}
\end{figure}

We design \cite {bormio,hybrid,hadron1} the COMPASS Primakoff trigger to enhance
the acceptance and statistics. We minimize target break-up events via veto
scintillators around the target. The trigger uses the characteristic decay
pattern: one or three charged mesons with accompanying $\gamma$ hits, or three
charged mesons and no $\gamma$ hits. The trigger \cite {bormio,hybrid,hadron1}
for the $\pi\eta$ hybrid decay channel (charged particle multiplicity =1) is
based on a determination of the pion energy loss (via its characteristic angular
deflection), correlated with downstream scintillator hodoscopes stations (H1
versus H2) with the aid of a fast matrix chip, as shown in
Fig.~\ref{fig:trigger}. This trigger is a copy of the currently running muon-beam
energy-loss trigger \cite {future}. We will use the Beam Kill detectors BK1/BK2
as veto only during low intensity tests. These detectors are positioned in the
pion-beam trajectory, as shown in Fig.~\ref{fig:trigger}, but they cannot handle
the full 100 MHz beam rate.

\section {OBJECTIVES AND EXPECTED SIGNIFICANCE}

COMPASS can study pionic hybrid-meson candidates between near 1.4-1.9 GeV,
produced by the ultra-peripheral Primakoff reaction. But COMPASS may also be
sensitive to pionic and kaonic hybrids for the 1.9-3.0 GeV mass range, if they
also couple to the $\pi \rho$ and $K \rho$ channels. We can then potentially
obtain superior statistics for hybrid states via a production mechanism that is
not complicated by hadronic final-state interactions. We can also get important
data on the different decay modes in both pionic and Kaonic channels.

We make initial rough estimates of the statistics attainable for hybrid
production in the COMPASS experiment. Monte Carlo simulations will refine these
estimates. We assume a 1.6 mb cross section per Pb nucleus for production of a
1.6 GeV Hybrid meson. We use the radiative width of $\Gamma(\pi 1(1600)
\rightarrow \pi \gamma) \approx 1000 keV$, as described above, and also for the
1.9-3.0 GeV mass region. Integrating the Primakoff Hybrid production differential
cross section for a 280 GeV pion beam, and using this radiative width, gives 1.6
mb \cite{hybrid,zihy,tf}.

 We consider a beam flux of $2 \times 10^{7}$ pions/sec, with a spill structure
that provides a 5 second beam every 16 seconds. In 2 months of running at 100\%
efficiency, we obtain 3.2$\times$ 10$^{13}$ beam pions. Prior to the data
production run, time is also needed to calibrate ECAL2, to make the tracking
detectors operational, to bring the DAQ to a stable mode, and for other
contingencies. We use a 0.8~\% interaction length target, or 1.4 mm Lead plate
with target density $N_t=10^{22} cm^{-2}$. The Primakoff interaction rate is then
$R = \sigma(Pb) \cdot N_t= 1.6 \times$ 10$^{-5}$. In a 2 month run period, we
therefore obtain 5.1$\times$ 10$^{8}$ Hybrid events at 100\% efficiency.
Considering efficiencies for tracking (92\%), $\gamma$ detection (58\% for each
$\gamma$), accelerator and COMPASS operation (70\%), analysis cuts to reduce
backgrounds (75\%), branching ratio for the $\eta \rightarrow 2 \gamma$ and $\eta
\rightarrow \pi^+ \pi^- \pi^0$ decay modes ($\approx 62$\%), trigger efficiency
($\approx 60$\%), geometrical acceptances ($\approx 90$\%), $\pi^0$ and $\eta$
reconstruction from two $\gamma$ hits ($\approx 45$\%), and event reconstruction
efficiencies ($\approx 60$\%), we estimate a global efficiency of
$\epsilon$(total)= 1.5\%. We will assume here the same average detection
efficiency for all Hybrid decay modes, via $\rho\pi, \eta\pi, \eta'\pi, \pi b_1,
\pi f_1$. Therefore, we can expect to observe a total of 7.7 $\times 10^{6}$
Hybrid decays in all decay channels for the 1.6 GeV Hybrid. For example,
following theory and VES branching ratios, we expect for the 1.6 GeV state, most
data in the $\pi f_1$, $\pi \rho$, $\pi \eta'$, and b$_1\pi$ channels.

For 2, 2.5, 3.0 GeV mass Hybrids, assuming that they have the same radiative
width, the number of useful events decreases by factors of 6, 25, and 100,
respectively. But even in these cases, assuming again a global 1.5\% efficiency,
that would represent a very interesting potential of samples of
13$\times$10$^{5}$, 3.1$\times$10$^{5}$, and 0.77 $\times 10^{5}$ Hybrid meson
detected events, with masses 2.0, 2.5, and 3.0 GeV, respectively.

Taking into account the very high beam intensity, fast data acquisition, high
acceptance and good resolution of the COMPASS setup, one can expect from COMPASS
the highest statistics and a `systematics-free' data sample that includes many
tests to control possible uncertainties. Comparison between COMPASS and past and
new experiments \cite {GSI,chpc}, with complementary methodologies, should allow
fast progress on understanding Hybrid meson structure, their production and decay
characteristics, and on establishing systematic uncertainties.

\vskip1cm
\noindent

\section*{ACKNOWLEDGEMENTS}

This work was supported in part by the Israel Science Foundation founded by the
Israel Academy of Sciences and Humanities. Thanks are due to T. Ferbel, V.
Dorofeev, and S. Paul for a critical reading of this manuscript. Thanks are due
also to R. Bertini, F. Bradamante, A. Bravar, D. Casey, S. U. Chung, M.
Colantoni, N. d'Hose, S. Donskov, W. Dunnweber, M. Faessler, M. Finger, L.
Frankfurt, S. Godfrey, H. Hahn, D. von Harrach, T. Hasegawa, Y. Khokhlov, K.
Koenigsmann, R. Kuhn, F. Kunne, L. Landsberg, J. Lichtenstadt, A. Magnon, G.
Mallot, V. Molchanov, J. Nassalski, A. Olchevski, E. Piasetzky, J. Pochodzalla,
S. Prakhov, A. Sandacz, L. Schmitt, C. Schwarz, H. W. Siebert, V. Sougoniaev, T.
Walcher, and M. Zielinski for valuable discussions. Thanks are due to the
Johannes Gutenberg Universitaet (Mainz, Germany) for hospitality during the
writing of this report, during a sabbatical leave from Tel Aviv University, as a
visiting Mercator Professor.


\end{document}